\documentclass[11pt]{article}

\usepackage[preprint]{acl}

\usepackage{times}
\usepackage{latexsym}
\usepackage{float}
\usepackage[T1]{fontenc}

\usepackage[utf8]{inputenc}
\usepackage{multirow}
\usepackage{microtype}

\usepackage{inconsolata}

\usepackage{graphicx}

\usepackage{booktabs}
\usepackage{graphicx}
\usepackage{multirow}
\usepackage{amsmath}
\usepackage{subcaption}
\usepackage{pifont}

\title{X-Translator: A Real-Time Multilingual Speaker-Aware Speech-to-Speech Translation System}

\author{
 \textbf{Yuxiang Zhao\textsuperscript{1,2}\thanks{These authors contributed equally to this work.}},
 \textbf{Yichi Zhang\textsuperscript{1}\footnotemark[1]},
 \textbf{Yanjie An\textsuperscript{1}\footnotemark[1]},
 \textbf{Yanqiao Zhu\textsuperscript{1,2}},
 \textbf{Zhanxun Liu\textsuperscript{1,2}},
 \textbf{Yushen Chen\textsuperscript{1,2}},
 \\
 \textbf{Qixi Zheng\textsuperscript{1,2}},
 \textbf{Haina Zhu\textsuperscript{1,2}},
 \textbf{Yunchong Xiao\textsuperscript{1}},
 \textbf{Keqi Deng\textsuperscript{3}},
 \textbf{Shuai Fan}\textsuperscript{4},
 \textbf{Kai Yu\textsuperscript{1}},
 \textbf{Xie Chen\textsuperscript{1,2}\thanks{Corresponding author.}},
\\
\\
 \textsuperscript{1}MoE Key Lab of Artificial Intelligence, Jiangsu Key Lab of Language Computing,\\
X-LANCE Lab, School of Computer Science, Shanghai Jiao Tong University\\
 \textsuperscript{2}Shanghai Innovation Institute
 \textsuperscript{3}Microsoft
 \textsuperscript{4}AISpeech Co., Ltd.
\\
 \small{
   \textbf{Correspondence:} {yuxiangzhao@sjtu.edu.cn, chenxie95@sjtu.edu.cn}
 }
}

\begin{document}
\maketitle
\begin{abstract}
Real-time speech-to-speech translation (S2ST) systems must balance translation quality, latency, speech naturalness, and speaker consistency. Publicly documented S2ST systems have advanced direct, multilingual, streaming, and expressive modeling, while proprietary products and APIs increasingly expose real-time translation capabilities to users. However, practical deployment remains challenging for open and reproducible systems, especially in long-form and multi-speaker conversations where partial ASR hypotheses are unstable, turn boundaries are ambiguous, and target speech must be generated with an appropriate speaker prompt. We present X-Translator, a low-cost modular cascaded S2ST system that combines streaming ASR, machine translation, and prompt-conditioned TTS through a session-level runtime controller. The system uses incremental segment commitment to convert unstable ASR streams into translation-ready units, and an online speaker prompt manager to bind source speech spans to speaker-specific voice prompts for synthesis. We evaluate translation, speech quality, and latency with OpenSTBench, compare against proprietary speech translation APIs as behavioral baselines, measure long-form voice stability, evaluate speaker preservation in multi-speaker conversations, and assess multilingual translation quality. X-Translator provides an open platform for understanding the practical trade-offs of deployment-oriented S2ST. Code and demo are available at \url{https://github.com/zhaoyx239/X-Translator}.
\end{abstract}
\section{Introduction}

Speech-to-speech translation (S2ST) aims to convert speech in one language into speech in another language while preserving the communicative intent of the original speaker. Compared with speech-to-text translation, S2ST better matches real-time cross-lingual interaction because users can listen and respond without reading intermediate transcripts. A deployable S2ST system, however, must optimize several objectives at once: translation quality, low latency, intelligible and natural target speech, and speaker consistency when multiple speakers appear in the same audio stream.

Publicly documented S2ST research systems have made rapid progress. Direct systems such as Translatotron and Translatotron~2 show that source speech can be mapped to translated speech without an explicit text interface, including early mechanisms for voice preservation~\cite{jia2019translatotron,jia2021translatotron2}. Large-scale systems such as SeamlessM4T and the Seamless family~\cite{barrault2023seamlessm4t,communication2023seamless} further unify multilingual speech recognition, speech translation, and speech generation, while recent systems such as StreamSpeech~\cite{zhang2024streamspeech}, UniSS~\cite{cheng2025uniss}, and Seed LiveInterpret~\cite{cheng2025seedliveinterpret} move toward simultaneous, expressive, or voice-preserving S2ST. These systems establish strong model-level capabilities. Complementarily, we study the runtime control needed to assemble replaceable components into a long-form, multi-speaker translation service and to make its decisions observable.

At the same time, proprietary speech translation products and APIs suggest that real-time S2ST is becoming a practical application. Systems such as Doubao AST 2.0\footnote{\url{https://www.volcengine.com/docs/4640/127504}}, Qwen3-LiveTranslate\footnote{\url{https://www.alibabacloud.com/help/en/model-studio/qwen3-livetranslate-flash-realtime}}, Qwen3.5-LiveTranslate\footnote{\url{https://www.alibabacloud.com/help/en/model-studio/qwen3-5-livetranslate-flash-realtime}}, and GPT Realtime Translate\footnote{\url{https://developers.openai.com/api/docs/models/gpt-realtime-translate}} have demonstrated strong user-facing translation capabilities in real-world scenarios. Yet these systems are largely black boxes: their model architectures, training data, streaming policies, and speaker-handling mechanisms are not publicly available. As a result, while they highlight the practical potential of real-time S2ST, their closed nature limits reproducible research and makes it difficult to inspect, adapt, or diagnose their behavior at the component level.


This paper presents X-Translator, an open and reproducible cascaded system for real-time S2ST. X-Translator composes replaceable automatic speech recognition (ASR), machine translation (MT), and prompt-conditioned text-to-speech (TTS) modules through a session-level runtime controller, allowing the system to inherit progress from independently improved speech and language components. The system converts unstable streaming ASR hypotheses into committed translation units, routes each unit to a speaker-specific voice prompt, and synthesizes target speech with speaker-aware cloning under a low-latency deployment setting.

By selecting lightweight backends, the modular architecture can be deployed on a single consumer-grade NVIDIA RTX 3090 GPU. At the same time, each component can be replaced as stronger or more suitable ASR, MT, and TTS models become available. The resulting system serves as a practical and reproducible real-time S2ST system for studying cascaded simultaneous speech translation, and its modular logs allow errors and delays to be traced to ASR commitment, MT, TTS, speaker routing, or frontend playback.

Evaluation is another bottleneck for deployment-oriented S2ST. Standard short-form benchmarks capture important aspects of translation and speech quality, but they do not fully characterize long-duration simultaneous interpretation, same-speaker voice consistency during streaming translation, or robust handling of multi-speaker turn switches. We therefore use OpenSTBench~\cite{an2026openstbench} as the main short-form evaluation protocol, and complement it with long-form, multi-speaker, and multilingual experiments that expose runtime stability, cross-lingual generalization, and speaker-aware voice consistency.

In summary, this paper makes the following contributions:
\begin{itemize}
    \item We release X-Translator, a fully open, reproducible, and modular cascaded S2ST system for real-time simultaneous translation. Its replaceable ASR, MT, and TTS modules allow the system to be adapted to different deployment requirements.
    \item We design an incremental segment-commitment layer that converts unstable streaming ASR hypotheses into translation-ready units without sending display-only partial hypotheses to downstream modules.
    \item We introduce an online speaker prompt manager that maintains speaker-specific prompt state and routes each committed segment to an appropriate prompt for synthesis.
    \item We contribute a deployment-oriented evaluation that complements short-form benchmarks with single-speaker long-form and multi-speaker long-form settings, alongside multilingual evaluation, to measure behavior not visible in isolated utterances.
\end{itemize}

\section{Related Work}

\subsection{Publicly Documented S2ST Systems}

Speech-to-speech translation has been studied through both cascaded and direct modeling. Cascaded systems compose ASR, MT, and TTS, which makes them modular and easy to inspect, but also exposes them to error propagation and accumulated latency. Direct systems attempt to map source speech to target speech more directly; Translatotron first demonstrated direct S2ST with a sequence-to-sequence architecture, and Translatotron~2 improved quality and voice preservation with a more structured design~\cite{jia2019translatotron,jia2021translatotron2}.

A related line of direct S2ST predicts discrete speech units instead of target spectrogram features. Lee et al. train a speech-to-unit translation (S2UT) model to predict discrete representations of the target speech, showing better performance than a direct spectrogram-prediction baseline and potential use in settings with limited text supervision~\cite{lee2022directs2ut}. Textless S2ST further studied translation without relying on paired text transcripts, while TranSpeech and DASpeech explored faster or non-autoregressive generation to reduce inference latency~\cite{lee2022textless,huang2023transpeech,fang2023daspeech}.

Large-scale multilingual systems broaden the scope of S2ST beyond single language pairs. CVSS provides a massively multilingual corpus with canonical-voice and transferred-voice settings for S2ST research~\cite{jia2022cvss}. SeamlessM4T unifies ASR, speech-to-text translation, text-to-speech translation, and S2ST across many languages, and the Seamless family extends this direction toward expressive and streaming speech translation~\cite{barrault2023seamlessm4t,communication2023seamless}.

Recent research also targets simultaneous, expressive, and voice-preserving S2ST. StreamSpeech jointly models translation and simultaneous policies for streaming target speech~\cite{zhang2024streamspeech}. UniSS studies a unified expressive S2ST framework that preserves voice and emotional style, while Seed LiveInterpret targets end-to-end simultaneous S2ST with the source speaker's voice~\cite{cheng2025uniss,cheng2025seedliveinterpret}. Our work is complementary: rather than proposing a new unified S2ST model, X-Translator studies a modular cascade in which the runtime policies are explicit, replaceable, and measurable.

\subsection{Proprietary Speech Translation Systems}

Commercial speech translation products and APIs are increasingly relevant as baselines because they represent systems that users can directly access. In this paper we treat Doubao, Qwen LiveTranslate, and GPT Realtime Translate as proprietary behavioral baselines rather than academic prior work. We compare only their observable behavior, including returned text or speech, latency, and stability.

The main limitation of proprietary baselines is opacity. Their model architecture, training data, streaming policy, endpoint segmentation, speaker handling, caching, and safety filters are usually unavailable to researchers. As a result, they can show what current product systems achieve, but they cannot reveal which internal decisions produce a failure or success. X-Translator is designed to complement these baselines by exposing the intermediate ASR hypotheses, committed source segments, translation requests, TTS prompts, and playback events that shape the final user experience.

\subsection{Streaming and Segment Commitment}

Streaming translation must decide when to emit target output before the complete source utterance is available. In simultaneous machine translation, prefix-based policies such as wait-$k$ explicitly model the latency-quality trade-off by generating target tokens after observing a limited source prefix~\cite{ma2019stacl}. Subsequent work has explored monotonic attention and learned policies for simultaneous translation~\cite{ma2019mma,liu2021caat,ma2021directsimuls2st}. Evaluation toolkits such as SimulEval formalize latency-aware evaluation for simultaneous text and speech translation~\cite{ma2020simuleval}.

Speech translation adds another layer of instability because streaming ASR hypotheses may change as more audio arrives. Prior SimulST work has studied chunk-based partial hypothesis selection, attention-guided read/write decisions, and segmented targets that better match simultaneous interpreting behavior~\cite{liu2020partialhypothesis,papi2023attention,papi2023alignatt,makinae2024simulmustc}. X-Translator shares the same latency-quality tension, but makes the decision at the system runtime level: it commits source-side ASR segments only when they are stable enough for downstream MT and TTS, turning a changing ASR stream into explicit translation units.
\begin{figure*}[t]
    \centering
    \includegraphics[width=\linewidth]{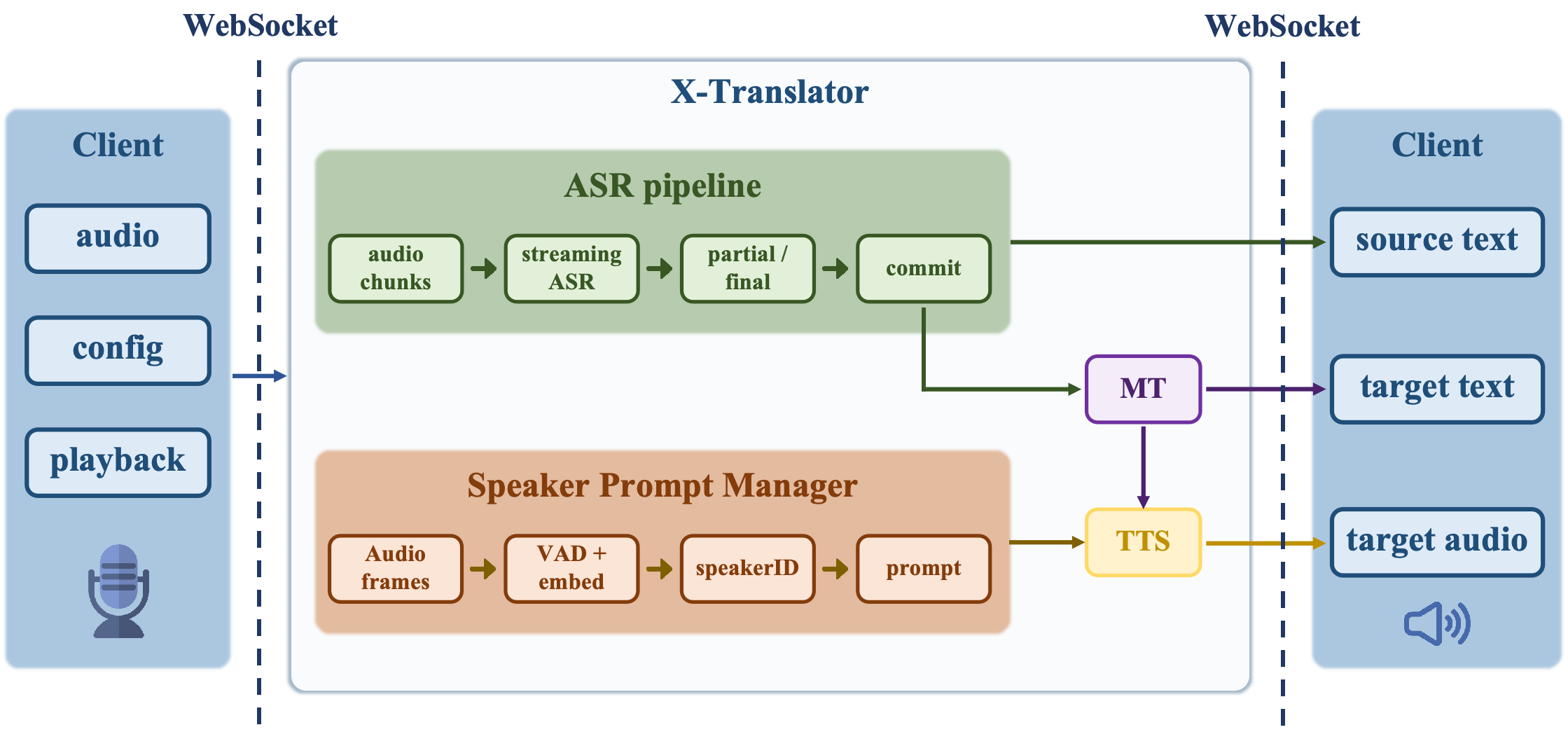}
    \caption{System architecture of X-Translator. The browser streams audio and session controls to the server through WebSocket. Inside X-Translator, the ASR pipeline commits source segments for display and translation, while the speaker prompt manager tracks speaker identity and routes the selected prompt to TTS. MT and TTS are shown as separate stages: MT produces target text, and TTS uses the selected prompt to synthesize target audio for browser playback.}
    \label{fig:system-architecture}
\end{figure*}
\subsection{Voice Preservation and Speaker-Aware Synthesis}

Voice preservation is central to S2ST because target speech should not only carry translated content, but also preserve speaker-related cues when possible. Direct S2ST systems have incorporated explicit voice-preservation mechanisms, and multilingual corpora such as CVSS include transferred-voice variants for studying this problem~\cite{jia2021translatotron2,jia2022cvss}. In parallel, prompt-conditioned and zero-shot TTS models have made it possible to synthesize speech from short acoustic prompts, as shown by neural codec and text-guided generative TTS systems~\cite{wang2023valle,le2023voicebox,du2024cosyvoice,chen-etal-2025-f5}.

These advances make speaker-aware S2ST increasingly feasible, but long-form multi-speaker input creates an additional runtime problem: the system must decide which source speaker each translated segment belongs to and which prompt should condition the target speech. Product-oriented and real-time S2ST systems have identified voice cloning, multi-speaker confusion, and speech inflation as deployment concerns~\cite{communication2023seamless,cheng2025seedliveinterpret}. X-Translator instantiates this design through a speaker prompt manager that maintains speaker-specific prompt buffers and routes each committed source segment to a selected prompt before synthesis.

\subsection{Evaluation for Speech Translation Systems}

Speech translation evaluation has expanded beyond text translation quality. Datasets such as CoVoST~2~\cite{wang2021covost2}, CVSS~\cite{jia2022cvss}, FLEURS~\cite{conneau2022fleurs}, and SpeechMatrix~\cite{duquenne2022speechmatrix} support multilingual speech translation research at different scales and modalities. However, S2ST system comparison remains difficult because outputs may differ in modality, timing, target speech realization, and speaker preservation. OpenSTBench addresses this by organizing heterogeneous speech translation outputs into a shared protocol that jointly evaluates translation quality, speech quality, temporal quality, latency, speaker preservation, and paralinguistic fidelity~\cite{an2026openstbench}.

OpenSTBench is useful for short-form system comparison, but deployment-oriented S2ST also requires tests that expose long-form streaming failures. A system may perform well on isolated utterances while still producing repeated segments, dropped content, accumulated latency, unstable partial outputs, or incorrect speaker prompts during a long session. We therefore use OpenSTBench for short-form evaluation and extend the analysis to long-form and multi-speaker scenarios where X-Translator's runtime logs make these failure modes observable.

\section{X-Translator}
\begin{figure*}[t]
    \centering
    \includegraphics[width=\textwidth]{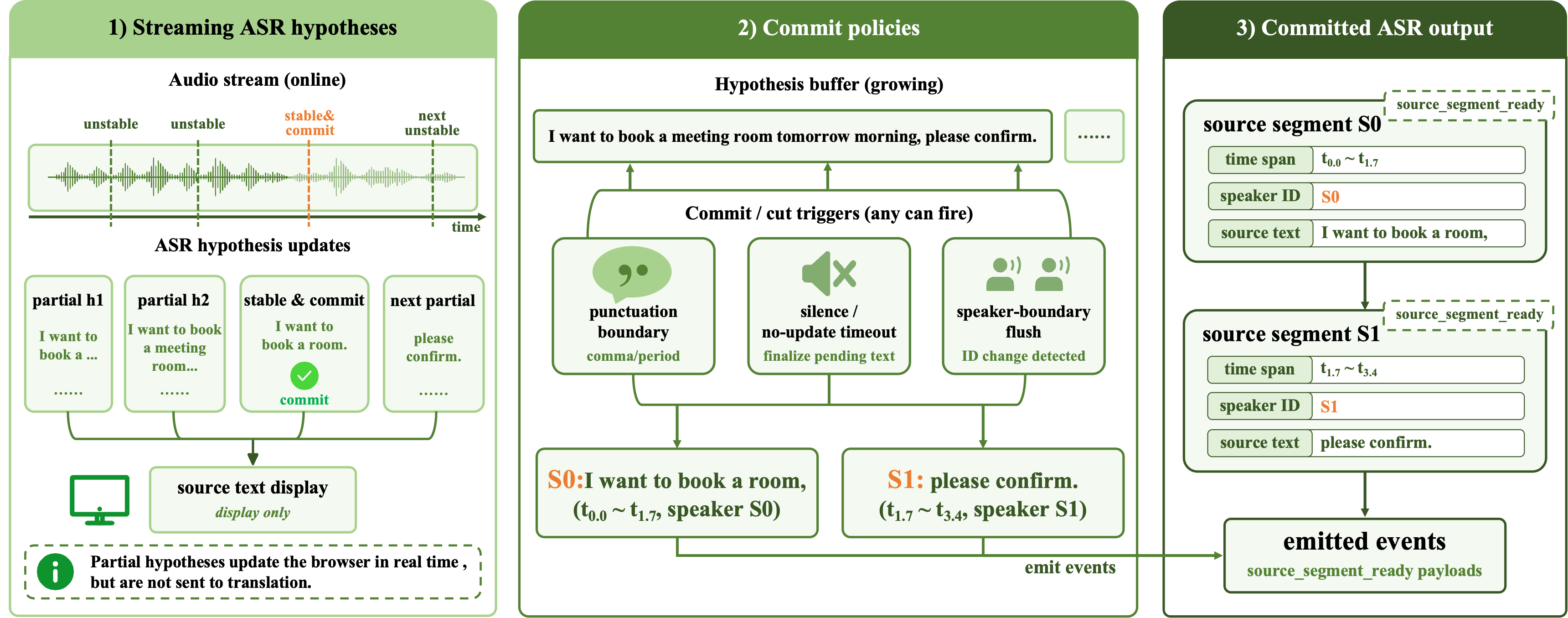}
    \caption{ASR pipeline and segment commitment policy. The left panel shows that streaming ASR hypotheses are continuously exposed to the browser for live source-text display, but partial hypotheses remain display-only and are not sent to MT or TTS. The middle panel shows the commitment layer, where any of several runtime triggers can cut a stable prefix from the growing hypothesis buffer: punctuation boundaries, silence or no-update timeouts, and speaker-boundary flushes. The right panel shows the committed ASR output emitted by the system. Each committed source segment is represented as a structured event containing a time span, speaker identifier, and source text, which becomes the unit consumed by downstream translation and synthesis.}
    \label{fig:asr-commit-policies}
\end{figure*}
\subsection{System Overview}

X-Translator is a modular cascaded S2ST system designed for real-time, long-form, and multi-speaker translation. The system receives streaming source audio from a browser client, processes it through a server-side session controller, and returns translated target speech for playback. Instead of treating speech translation as a single opaque model call, X-Translator exposes the main runtime decisions that determine the user experience: when a partial ASR hypothesis becomes stable enough to translate, which speaker a committed segment belongs to, which prompt should condition synthesis, and how much latency each module contributes.

Figure~\ref{fig:system-architecture} shows the system architecture. The browser streams audio frames and session controls to the server through WebSocket. The server maintains a session state that contains the ASR hypothesis buffer, committed source segments, translation requests, TTS requests, speaker prompt buffers, and playback metadata. The ASR pipeline continuously emits partial source text for live display, but only committed source segments are sent to MT and TTS. In parallel, the speaker prompt manager tracks speaker activity and provides a prompt audio clip for each committed segment before synthesis.

\subsection{Incremental Segment Commitment}

Streaming ASR produces hypotheses that are useful for immediate feedback but unsafe for downstream translation. A partial hypothesis may be extended, revised, or deleted as more audio arrives, and sending every partial hypothesis to MT and TTS would cause repeated translations and speech inflation. X-Translator therefore separates display from commitment: partial hypotheses can be shown to the user in real time, while the downstream pipeline only consumes committed source segments.

The commitment layer maintains a growing source-text buffer and cuts stable prefixes using several runtime triggers. A punctuation trigger commits a prefix when the ASR output reaches a strong textual boundary. A silence or no-update trigger commits buffered text when the speaker pauses or the ASR stream stops changing for a configured interval. A speaker-boundary trigger flushes the current buffer when the speaker prompt manager detects a speaker change, preventing a single translated segment from mixing two speakers. Each committed segment is emitted as a structured event with source text, start time, end time, speaker identifier, and diagnostic metadata.

This design makes the latency-quality trade-off explicit. Aggressive commitment can reduce waiting time, but may send incomplete context to MT and produce unnatural target speech. Conservative commitment can improve source completeness, but increases latency and may delay the first audible target output. Because X-Translator records both partial hypotheses and committed events, the system can later analyze whether an error came from ASR instability, premature commitment, MT, TTS, or playback scheduling.

\subsection{Speaker-Aware Prompt Management}
\label{sec:speaker-prompt-management}

Prompt-conditioned TTS makes it possible to synthesize target speech with speaker-specific acoustic cues, but it creates a routing problem in multi-speaker input. The system must decide which source speaker owns each committed segment and must maintain prompt audio that is clean, recent, and long enough to condition the TTS backend. X-Translator addresses this with an online speaker prompt manager that runs alongside ASR and segment commitment.

\begin{figure*}[t]
    \centering
    \includegraphics[width=\textwidth]{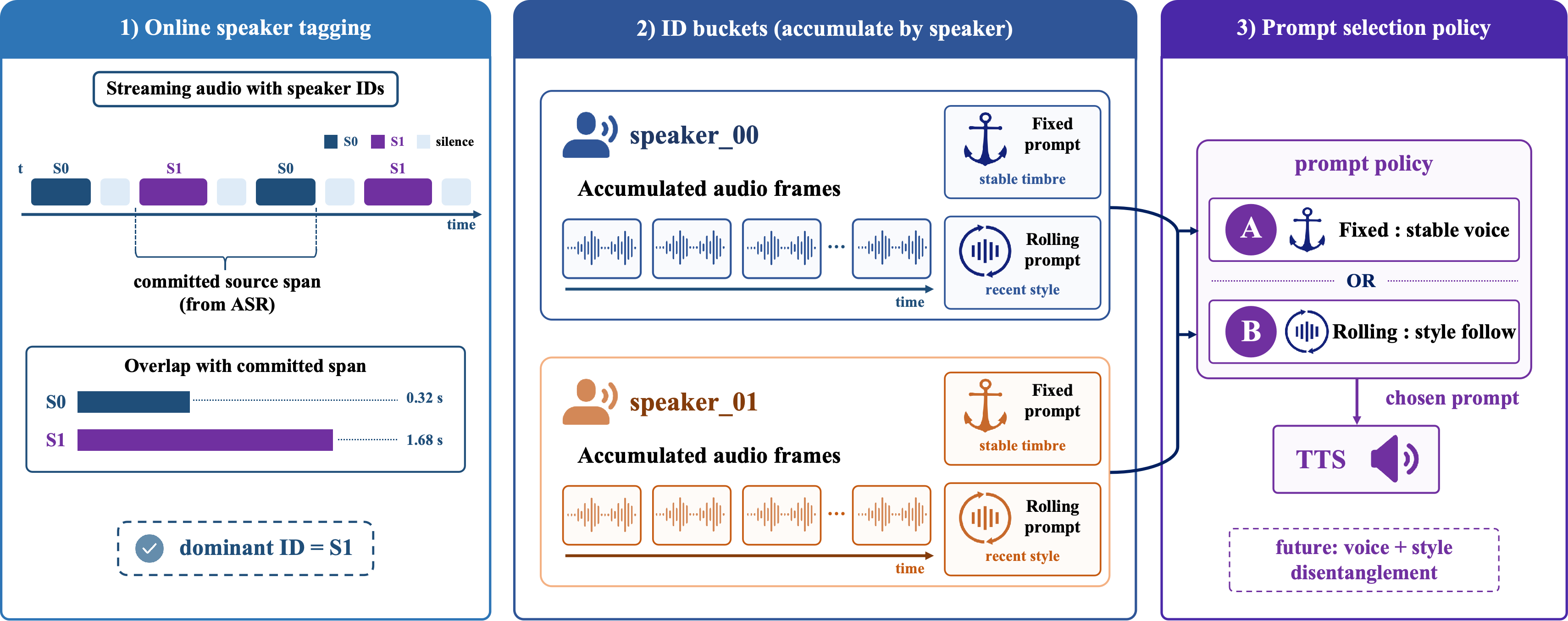}
    \caption{Speaker prompt manager. The left panel shows online speaker tagging over the streaming audio timeline. When the ASR pipeline emits a committed source span, the prompt manager assigns the segment to the speaker with the largest temporal overlap. The middle panel shows the per-speaker prompt buckets: each speaker independently maintains a fixed prompt from the first six seconds of speech and a rolling prompt from the most recent six seconds. The right panel shows the prompt selection policy used before TTS. Depending on prompt availability and the desired synthesis behavior, the system routes either the fixed or rolling prompt of the selected speaker to the TTS backend.}
    \label{fig:speaker-prompt-manager}
\end{figure*}

The prompt manager tracks speech activity and speaker changes over the input timeline using CAM++~\cite{wang2023cam++} speaker embeddings. As illustrated in the left panel of Figure~\ref{fig:speaker-prompt-manager}, each online speaker ID is associated with one or more time intervals. When the ASR pipeline commits a source segment, its start and end timestamps define an ASR span on the same timeline. The manager computes the temporal overlap between this ASR span and every speaker interval, and assigns the segment to the speaker ID with the largest overlap. In the example, the committed span overlaps both S1 and S2, but the S1 overlap is longer; the source segment is therefore bound to S1 and later synthesized using S1's prompt state. This overlap rule also handles ASR segments whose boundaries do not exactly coincide with speaker-change boundaries.

After binding the segment to a speaker ID, the manager updates that speaker's prompt bucket, shown in the middle panel of Figure~\ref{fig:speaker-prompt-manager}. Each bucket maintains two six-second prompt views. The \emph{fixed prompt} consists of the first six seconds of usable speech accumulated for that speaker and remains unchanged once initialized, providing a stable voice reference throughout the session. The \emph{rolling prompt} consists of the most recent six seconds of usable speech from the same speaker and is updated as new speech arrives, allowing the prompt to follow more recent vocal characteristics. Prompt audio is maintained independently for each speaker, so turns from one speaker do not overwrite another speaker's reference.

Before TTS, X-Translator selects the fixed or rolling six-second prompt from the bucket associated with the bound speaker ID, according to the configured synthesis mode. If a complete six-second prompt is not yet available, the system can use the available speaker audio or fall back to a default voice. This fallback behavior is important during the first turns of a conversation, before enough speaker-specific audio has accumulated. The resulting prompt and translated text are then sent together to the TTS backend, as shown in the right panel of Figure~\ref{fig:speaker-prompt-manager}.

\subsection{Translation and Speech Synthesis}

After a source segment is committed, X-Translator treats it as the atomic unit for downstream processing. The MT module receives stable source text, source-language and target-language identifiers, and optional session configuration, then returns target-language text. The TTS module receives the translated text and the selected speaker prompt, then generates target audio for browser playback. ASR, MT, and TTS are executed serially for each committed segment. This separation decouples the runtime question of when to translate from the model question of how to translate and synthesize.

The system only sends committed segments to MT and TTS. Temporary ASR text shown in the browser is display-only and never triggers target speech generation. This rule avoids possible repeated playback caused by ASR revisions and gives the session controller a single place to enforce ordering, cancellation, and queue management. If the TTS backend returns audio faster than playback can consume it, the browser queue preserves segment order; if a backend call fails, the system can mark the segment as failed without corrupting the ASR or prompt state.

The cascaded workflow also supports fine-grained latency measurement. X-Translator records ASR commit time, MT request and response time, TTS request and response time, and browser playback time. These logs decompose end-to-end delay into recognition waiting, translation cost, synthesis cost, and frontend playback scheduling, providing a common evidence trail for short-form, long-form, and multi-speaker evaluation.

\subsection{Runtime Modularity}

X-Translator organizes runtime logic as a session-level controller plus replaceable backends. The session controller maintains audio streams, ASR hypothesis buffers, committed segments, speaker prompt state, backend queues, and frontend events. ASR, MT, and TTS backends are connected through explicit input-output contracts: ASR returns incremental text with timing metadata, MT returns target text, and TTS returns target audio with synthesis metadata.

This modularity has two research benefits. First, it lets us compare backend models under the same streaming policy and evaluation protocol, separating the effect of runtime control from the capability of a particular ASR, MT, or TTS model. Second, it supports behavioral comparison with proprietary speech translation APIs: even when a commercial system is internally invisible, X-Translator can be analyzed through matched end-to-end outputs, latency, stability, and speaker consistency.

The modular design also introduces clear engineering trade-offs. A cascade may suffer from ASR error propagation, incompatible segmentation assumptions between modules, and accumulated network latency across services. We accept these trade-offs because the resulting system is interpretable, replaceable, and measurable. X-Translator is therefore both a usable S2ST prototype and a controlled platform for studying how segment commitment, prompt selection, and module latency jointly shape real-time S2ST behavior.

\section{Experiment Setup}

\subsection{Datasets}

We organize the evaluation into four parts: short-form benchmark evaluation, long-form streaming evaluation, multi-speaker evaluation, and multilingual evaluation. Short-form evaluation follows the OpenSTBench protocol under controlled utterance-level inputs. Long-form evaluation tests voice stability over continuous single-speaker audio. Multi-speaker evaluation is treated as a separate setting rather than a subset of long-form evaluation, because it focuses on speaker switching and prompt routing. Multilingual evaluation tests whether the cascade remains usable across language directions.

\textbf{Short-form.} For short-form evaluation, we follow OpenSTBench~\cite{an2026openstbench} rather than redefining the underlying benchmark datasets. We convert X-Translator and baseline outputs into the audio and text formats required by OpenSTBench, and record timing information from source input to playable target speech.

\textbf{Long-form.} For long-form single-speaker evaluation, we sample English audio from TEDLIUM-3~\cite{hernandez2018ted} and Chinese audio from WenetSpeech~\cite{zhang2022wenetspeech}. For each dataset, we randomly select ten clear recordings of at least ten minutes and simulate streaming input by feeding each recording to the system continuously. These samples are used to test whether translated speech preserves a stable target voice over extended sessions in English-to-Chinese and Chinese-to-English directions.

\textbf{Multi-speaker.} For multi-speaker evaluation, we sample long audio from the VoxConverse~\cite{voxconverse} dataset and a multi-speaker Chinese interview dataset collected from Bilibili. These samples are used to test whether the translated speech preserves a stable target voice and accurate speaker segmentation over extended sessions in Chinese-to-English and English-to-Chinese directions.

\textbf{Multilingual.} For multilingual evaluation, we use the FLEURS~\cite{conneau2022fleurs} dataset to assess speech translation across a broad range of languages. In addition to Chinese and English, the final evaluation includes 16 input languages and 19 output languages supported by the evaluated backend stack. These sets describe the configuration used in this study rather than a limitation of the X-Translator framework, whose modules can be replaced to support additional languages. Since the available utterances are not strictly aligned across all language subsets, for each translation direction, we evaluate only on samples with shared sentence IDs in both the source and target languages, ensuring semantically aligned and consistent test sets.


\subsection{Baselines}

For external comparison, we compare X-Translator against proprietary speech translation products or APIs, including Doubao, Qwen LiveTranslate, and GPT Realtime Translate. These baselines represent user-accessible black-box systems; because their internal segmentation, translation, synthesis, caching, and speaker-routing mechanisms are unavailable, the comparison focuses on observable outputs and timing behavior. The evaluations were collected at different times: the short-form OpenSTBench experiments use Qwen3-LiveTranslate, whereas the later long-form single- and multi-speaker experiments use its then-current successor, Qwen3.5-LiveTranslate. We retain the version actually evaluated in each table rather than treating results from the two service versions as interchangeable.


\subsection{Implementation Configuration}

Unless otherwise stated, all X-Translator experiments use Qwen3-ASR~\cite{shi2026qwen3asr} for streaming speech recognition, LMT-60-8B~\cite{luo2025lmt} for machine translation, and X-Voice~\cite{xu2026xvoice} for prompt-conditioned speech synthesis. We keep this backend stack fixed across the reported evaluations so that differences in system behavior can be attributed to the runtime policies and evaluation settings rather than to changes in component models. The modular implementation also supports replacing these backends; Table~\ref{tab:supported-backends} summarizes the currently supported models and their roles.

\begin{table*}[t]
  \centering
  \caption{Backend models supported by X-Translator.}
  \label{tab:supported-backends}
  \footnotesize
  \begin{tabular}{lllp{0.38\textwidth}}
    \toprule
    Module & Supported model & Parameter count & Role in X-Translator \\
    \midrule
    \multirow{4}{*}{ASR} & Qwen3-ASR~\cite{shi2026qwen3asr} & 1.7B & Streaming source speech recognition \\
    & X-ASR\textsuperscript{\dag} & 160M & Streaming source speech recognition \\
    & Paraformer-zh~\cite{gao2022paraformer} & 220M & Source speech recognition \\
    & SenseVoice-Small~\cite{an2024sensevoice} & 234M & Source speech recognition \\
    \midrule
    \multirow{2}{*}{MT} & LMT-60-4B/8B~\cite{luo2025lmt} & 4B/8B & Source-to-target text translation \\
    & HY-MT1.5~\cite{zheng2025hymt15} & 7B & Source-to-target text translation \\
    \midrule
    \multirow{2}{*}{TTS} & IndexTTS~\cite{deng2025indextts,zhou2026indextts2} & $\sim$0.5B\textsuperscript{*} & Prompt-conditioned target speech synthesis \\
    & X-Voice~\cite{xu2026xvoice} & 0.4B & Prompt-conditioned target speech synthesis \\
    \bottomrule
  \end{tabular}
  \parbox{0.96\textwidth}{
  \vspace{2pt}\footnotesize
  \textsuperscript{\dag}\url{https://github.com/Gilgamesh-J/X-ASR}\\
  \textsuperscript{*}Approximate parameter count estimated from the publicly released model weights; the IndexTTS paper does not report the total parameter count of the complete system.
}
\end{table*}
\subsection{Evaluation Protocol}

\subsubsection{Short-form} 

Short-form evaluation follows the OpenSTBench metric suite. For the S2ST results, we transcribe the generated target speech and compute BLEU, chrF++, COMET, and BLEURT from this target-ASR transcript; these metrics therefore capture errors from translation, synthesis, and target-speech recognition. We separately report S2TT quality computed from the systems' directly returned translation text, before speech synthesis and target-ASR transcription. CER is used for Chinese target speech and WER for English target speech. Start denotes the first playable target-audio offset, and cATD denotes the custom ATD variant computed from our runtime logs. Spk. SIM and Emo. SIM measure speaker and emotion similarity between source and generated target speech; for Spk. SIM, we report the Resemblyzer-based version.

Because several proprietary API baselines showed unstable availability or variable response behavior during collection, we run each short-form baseline evaluation twice and report the mean of the two runs. For X-Translator, we additionally record runtime logs so that latency, commitment, and prompt-routing errors can be traced to system events.
\begin{table*}[t]
  \centering
  \caption{Short-form English--Chinese speech-to-speech translation results under the OpenSTBench protocol. Translation-quality metrics are computed after transcribing the generated target speech with ASR. Higher values are better for translation-quality and similarity metrics; lower values are better for error and latency metrics. Baseline values are averaged over two runs. \textbf{Bold} values indicate the best result and \underline{underlined} values indicate the second-best result in each metric.}
  \label{tab:short-form-openstbench}
  \footnotesize
  \setlength{\tabcolsep}{3pt}
  \textbf{(a) English to Chinese}
  \vspace{2pt}

  \resizebox{\textwidth}{!}{%
  \begin{tabular}{lccccccccccccc}
    \toprule
    System &
    \multicolumn{4}{c}{Translation Quality} &
    \multicolumn{4}{c}{Speech Quality} &
    \multicolumn{3}{c}{Latency} &
    \multicolumn{2}{c}{Temporal Consistency} \\
    \cmidrule(lr){2-5}
    \cmidrule(lr){6-9}
    \cmidrule(lr){10-12}
    \cmidrule(lr){13-14}
    &
    BLEU$\uparrow$ &
    chrF++$\uparrow$ &
    COMET$\uparrow$ &
    BLEURT$\uparrow$ &
    UTMOS$\uparrow$ &
    CER(\%)$\downarrow$ &
    Spk. SIM$\uparrow$ &
    Emo. SIM$\uparrow$ &
    Start (ms)$\downarrow$ &
    ATD (ms)$\downarrow$ &
    cATD (ms)$\downarrow$ &
    SLC$_{0.2}\uparrow$ &
    SLC$_{0.4}\uparrow$ \\
    \midrule
    Qwen3-LiveTranslate & \textbf{27.06} & \textbf{17.96} & \textbf{0.81} & \textbf{0.61} & \textbf{3.58} & \textbf{3.35} & 0.60 & 0.63 & 2337 & 4723 & 3479 & \textbf{0.44} & \textbf{0.78} \\
    Doubao AST 2.0 & \underline{24.54} & 16.72 & \underline{0.79} & \underline{0.58} & 2.83 & \underline{4.35} & \underline{0.84} & \textbf{0.69} & \underline{2248} & \underline{4053} & \underline{2981} & 0.27 & \underline{0.69} \\
    GPT Realtime Translate & 14.39 & 10.74 & 0.74 & 0.49 & \underline{3.29} & 9.66 & 0.69 & \underline{0.68} & \textbf{814} & \textbf{2381} & \textbf{1143} & \underline{0.36} & 0.58 \\
    \textbf{X-Translator (ours)} & 24.17 & \underline{16.85} & 0.78 & \underline{0.58} & 2.66 & 5.88 & \textbf{0.87} & 0.67 & 2584 & 6234 & 3804 & 0.31 & 0.57 \\
    \bottomrule
  \end{tabular}%
  }

  \vspace{6pt}
  \textbf{(b) Chinese to English}
  \vspace{2pt}

  \resizebox{\textwidth}{!}{%
  \begin{tabular}{lccccccccccccc}
    \toprule
    System &
    \multicolumn{4}{c}{Translation Quality} &
    \multicolumn{4}{c}{Speech Quality} &
    \multicolumn{3}{c}{Latency} &
    \multicolumn{2}{c}{Temporal Consistency} \\
    \cmidrule(lr){2-5}
    \cmidrule(lr){6-9}
    \cmidrule(lr){10-12}
    \cmidrule(lr){13-14}
    &
    BLEU$\uparrow$ &
    chrF++$\uparrow$ &
    COMET$\uparrow$ &
    BLEURT$\uparrow$ &
    UTMOS$\uparrow$ &
    WER(\%)$\downarrow$ &
    Spk. SIM$\uparrow$ &
    Emo. SIM$\uparrow$ &
    Start (ms)$\downarrow$ &
    ATD (ms)$\downarrow$ &
    cATD (ms)$\downarrow$ &
    SLC$_{0.2}\uparrow$ &
    SLC$_{0.4}\uparrow$ \\
    \midrule
    Qwen3-LiveTranslate & \textbf{22.26} & \textbf{48.62} & \underline{0.75} & \underline{0.61} & \textbf{4.20} & 6.91 & 0.57 & 0.69 & 3462 & 7966 & 6224 & 0.33 & 0.67 \\
    Doubao AST 2.0 & \underline{19.50} & 43.59 & \textbf{0.77} & \textbf{0.63} & 3.43 & \underline{3.70} & \underline{0.84} & \textbf{0.86} & \underline{3149} & \underline{6278} & \underline{5093} & 0.16 & 0.56 \\
    GPT Realtime Translate & 16.27 & 41.99 & 0.73 & 0.59 & \underline{3.92} & 8.30 & 0.66 & \underline{0.73} & \textbf{800} & \textbf{3677} & \textbf{2657} & \textbf{0.48} & \textbf{0.86} \\
    \textbf{X-Translator (ours)} & 15.91 & \underline{45.21} & 0.74 & 0.59 & 3.39 & \textbf{3.04} & \textbf{0.88} & \textbf{0.86} & 4056 & 9125 & 6566 & \underline{0.39} & \underline{0.70} \\
    \bottomrule
  \end{tabular}%
  }
  \vspace{2pt}
\end{table*}

\begin{table*}[t]
  \centering
  \caption{Short-form speech-to-text translation (S2TT) quality under the OpenSTBench protocol. Metrics are computed from each system's directly returned translation text, before speech synthesis or target-speech ASR. Baseline values are averaged over two runs. \textbf{Bold} values indicate the best result and \underline{underlined} values indicate the second-best result in each metric.}
  \label{tab:short-form-s2tt}
  \footnotesize
  \setlength{\tabcolsep}{6pt}
  \begin{tabular}{lcccccccc}
    \toprule
    & \multicolumn{4}{c}{English to Chinese} & \multicolumn{4}{c}{Chinese to English} \\
    \cmidrule(lr){2-5}\cmidrule(lr){6-9}
    System & BLEU$\uparrow$ & chrF++$\uparrow$ & COMET$\uparrow$ & BLEURT$\uparrow$ & BLEU$\uparrow$ & chrF++$\uparrow$ & COMET$\uparrow$ & BLEURT$\uparrow$ \\
    \midrule
    Qwen3-LiveTranslate & \textbf{42.72} & \textbf{29.45} & \textbf{0.85} & \textbf{0.68} & \textbf{24.30} & \textbf{50.46} & \textbf{0.78} & \textbf{0.63} \\
    Doubao AST 2.0 & \underline{36.89} & \underline{25.10} & \textbf{0.85} & \underline{0.67} & \underline{20.84} & 44.69 & \textbf{0.78} & \textbf{0.63} \\
    GPT Realtime Translate & 22.44 & 15.64 & 0.79 & 0.58 & 15.11 & 41.94 & 0.73 & 0.58 \\
    \textbf{X-Translator (ours)} & 32.44 & 24.19 & \underline{0.83} & 0.66 & 15.56 & \underline{45.22} & \underline{0.74} & \underline{0.60} \\
    \bottomrule
  \end{tabular}
\end{table*}


\subsubsection{Long-form} 

For the long-form single-speaker voice-stability test, we remove non-speech regions with VAD and normalize each source and target timeline by cumulative effective speech duration, so that source and translated speech can be compared despite duration mismatch. We segment each audio stream into 10-second windows with a 5-second hop, extract speaker embeddings using the Resemblyzer\footnote{\url{https://github.com/resemble-ai/Resemblyzer}} voice encoder, and compute cosine similarity between $L_2$-normalized embeddings. Source-Source Long-Gap Similarity (SSLG) measures the natural long-range consistency of the source speaker by averaging source-window pairs whose normalized progress gap is at least $\gamma$; we use $\gamma=0.5$. Target-Target Long-Gap Similarity (TTLG) applies the same computation to generated target speech, while Global Source Similarity (GSS) compares each target-window embedding with the normalized average source-speaker embedding.

\subsubsection{Multi-speaker.} 

For the long-form multi-speaker evaluation, we first obtain speaker-attributed transcripts for both the source and translated audio using VibeVoice-ASR\footnote{\url{https://huggingface.co/microsoft/VibeVoice-ASR}}~\cite{peng2026vibevoiceasr}. Each audio stream is represented as a sequence of speech segments with start time, end time, speaker label, and transcribed text. Non-speech and invalid segments are removed, and very short adjacent segments assigned to the same speaker are merged to reduce over-fragmentation.

To establish the alignment, we primarily perform semantic matching by computing the multilingual semantic similarity between the source and target ASR texts using DashScope text-embedding-v4 from Qwen3-Embedding~\cite{zhang2025qwen3embedding} series, selecting the target segment with the highest cosine similarity. However, relying solely on semantics can occasionally lead to distant misalignments in long-form audio. To mitigate such errors and restrict the search space, we introduce a local temporal constraint: candidate target segments are filtered to those whose midpoint percentile differs by at most $0.10$ from the source segment's midpoint percentile. This relative position normalization (percentile) is adopted instead of absolute time alignment, as the source and translated audio often differ in duration.

After semantic alignment, we crop the corresponding source and target audio spans using their own ASR-derived timestamps. Speaker embeddings are extracted from each matched audio pair using the Resemblyzer voice encoder, and cosine similarity between $L_2$-normalized embeddings is used as the segment-level voice similarity. The final multi-speaker voice similarity is the source-duration-weighted average over all matched segment pairs:
\begin{equation}
\text{Spk. SIM} = \frac{\sum_i d_i \cdot \cos(\mathbf{e}_{\text{src}, i}, \mathbf{e}_{\text{tgt}, i})}{\sum_i d_i}
\end{equation}
where $d_i$ is the duration of the $i$-th source segment, and $\mathbf{e}_{\text{src}, i}$ and $\mathbf{e}_{\text{tgt}, i}$ are the corresponding source and target speaker embeddings.

To evaluate speaker segmentation consistency, we construct a duration-weighted source-target speaker correspondence matrix $C$. For every semantically matched segment pair, we add the source segment duration to the cell corresponding to its source speaker $s$ and matched target speaker $t$:
\begin{equation}
C[s,t] \leftarrow C[s,t] + d_i
\end{equation}
We then compute target speaker purity as:
\begin{equation}
\text{TargetPurity} = \frac{\sum_t \max_s C[s,t]}{\sum_{s,t} C[s,t]}
\end{equation}
This metric measures whether each generated target speaker is dominated by a single source speaker. It penalizes speaker collapse, where speech from multiple source speakers is mapped to the same target speaker. To also measure speaker fragmentation, we compute source speaker purity in the reverse direction:
\begin{equation}
\text{SourcePurity} = \frac{\sum_s \max_t C[s,t]}{\sum_{s,t} C[s,t]}.
\end{equation}
SourcePurity is high when the translated speech associated with each source speaker is concentrated in one target-speaker cluster, and decreases when a source speaker is split across multiple target speakers. Target Purity and Source Purity should be interpreted jointly, since either metric alone may be misleading. In particular, a system that maps all source speakers to a single fixed target voice can still obtain a Source Purity close to 1, despite having no effective speaker-cloning capability. We report $\text{Spk. SIM}$, $\text{TargetPurity}$, and $\text{SourcePurity}$ for the multi-speaker setting.

\subsubsection{Multilingual.}

We evaluate multilingual S2ST performance using reference-based COMET~\footnote{\url{https://huggingface.co/Unbabel/wmt22-comet-da}}. For each language direction, we first transcribe the generated target speech using a target-language ASR model and then compute COMET between the resulting transcription and the reference translation. Thus, the reported multilingual COMET is an ASR-based final-speech metric, not a score computed from the MT module's directly returned text; it reflects errors introduced by translation and synthesis as well as target-speech ASR. We compute COMET separately for each direction and report macro-averaged scores over the corresponding direction groups.

\section{Results}

\subsection{Short-Form Speech Translation}

Table~\ref{tab:short-form-openstbench} evaluates the complete S2ST output: its translation-quality metrics are computed from an ASR transcription of the generated target speech. No system dominates all dimensions. X-Translator's clearest advantage is speaker preservation: it achieves the highest Spk. SIM in both English-to-Chinese (0.87) and Chinese-to-English (0.88). In Chinese-to-English, it also obtains the lowest target-speech WER (3.04\%) and ties Doubao AST 2.0 for the highest emotion similarity (0.86). These results indicate that the prompt-conditioned cascade retains speaker characteristics effectively and, particularly for English output, produces speech whose content can be recovered reliably by ASR.

This preservation comes with a quality--latency trade-off. In the ASR-based S2ST evaluation, X-Translator reaches COMET scores of 0.78 and 0.74. For English-to-Chinese it trails Qwen3-LiveTranslate (0.81) and Doubao AST 2.0 (0.79); for Chinese-to-English, Doubao obtains the highest COMET (0.77), followed by Qwen3-LiveTranslate (0.75). X-Translator is also the slowest system in both directions: its start offsets are 2,584 and 4,056~ms, compared with 814 and 800~ms for GPT Realtime Translate, and the same gap is visible in ATD and cATD. GPT Realtime Translate occupies the opposite end of this trade-off, giving the fastest responses but weaker translation scores and higher target-speech error rates.

Table~\ref{tab:short-form-s2tt} isolates translation before speech synthesis by scoring each system's directly returned text. X-Translator's S2TT COMET rises to 0.83 for English-to-Chinese, compared with 0.78 after synthesis and target ASR, while its Chinese-to-English COMET remains 0.74 under both protocols. The difference between the two tables should not be attributed to MT alone: the S2ST scores additionally include errors introduced by synthesis and by the evaluation ASR. Under S2TT, Qwen3-LiveTranslate and Doubao AST 2.0 tie for the highest COMET in both directions (0.85 and 0.78, respectively), while Qwen3-LiveTranslate leads the other translation metrics overall.

The success rates in Table~\ref{tab:openstbench-success-rate} provide an important qualification: the quality metrics are computed over successful outputs, while unsuccessful or semantically incomplete calls are reflected separately in coverage. X-Translator succeeds on 98.92--100\% of the evaluated subsets; the proprietary baselines range from 93.50\% to 100\%. Thus, the short-form results support X-Translator primarily as a high-coverage, speaker-preserving system, rather than as the best system for either translation quality or latency.

\subsection{Long-Form Single-Speaker Voice Stability}

\begin{table}[t]
  \centering
  \caption{Evaluation results of long-form speech-to-speech translation voice stability on single-speaker long audio. \textbf{SSLG} measures Source-Source Long-Gap similarity, \textbf{TTLG} measures Target-Target Long-Gap similarity, and \textbf{GSS} measures Global Source Similarity.}
  \label{tab:voice-stability-simplified}
  \footnotesize
  \textbf{(a) English to Chinese (TEDLIUM-3)}
  \vspace{2pt}

  \begin{tabular}{lccc}
    \toprule
    System & SSLG & TTLG $\uparrow$ & GSS $\uparrow$ \\
    \midrule
    Qwen3.5-LiveTranslate & 0.89 & 0.94 & 0.53 \\
    Doubao AST 2.0 & 0.89 & 0.92 & 0.83 \\
    GPT Realtime Translate & 0.89 & 0.84 & 0.69 \\
    \textbf{X-Translator (ours)} & 0.89 & 0.91 & 0.83 \\
    \bottomrule
  \end{tabular}

  \vspace{6pt}
  \textbf{(b) Chinese to English (WenetSpeech)}
  \vspace{2pt}

  \begin{tabular}{lccc}
    \toprule
    System & SSLG & TTLG $\uparrow$ & GSS $\uparrow$ \\
    \midrule
    Qwen3.5-LiveTranslate & 0.89 & 0.94 & 0.50 \\
    Doubao AST 2.0 & 0.89 & 0.90 & 0.79 \\
    GPT Realtime Translate & 0.89 & 0.84 & 0.66 \\
    \textbf{X-Translator (ours)} & 0.89 & 0.90 & 0.79 \\
    \bottomrule
  \end{tabular}
\end{table}

Table~\ref{tab:voice-stability-simplified} separates target-side consistency (TTLG) from faithfulness to the source voice (GSS). X-Translator obtains TTLG scores of 0.91 for English-to-Chinese and 0.90 for Chinese-to-English, close to the source recordings' SSLG of 0.89. Its synthesized voice therefore remains internally consistent across distant portions of a continuous session, with no evidence of substantial long-range drift under this metric.

High target-side consistency alone does not guarantee source-speaker preservation. Qwen3.5-LiveTranslate records the highest TTLG in both directions (0.94), but its GSS is only 0.53 and 0.50. By contrast, X-Translator reaches GSS scores of 0.83 and 0.79, matching Doubao AST 2.0 and exceeding the other two baselines. Doubao is slightly more internally consistent for English-to-Chinese (0.92 versus 0.91) and ties X-Translator for Chinese-to-English (0.90). Taken together, TTLG and GSS show that X-Translator maintains a stable target voice while preserving substantially more source-speaker identity than a system that produces a highly consistent but less source-matched voice.

\subsection{Long-Form Multi-Speaker Voice Preservation}

\begin{table*}[t]
  \centering
  \caption{Long-form multi-speaker speech-to-speech translation results. Spk. SIM measures speaker identity preservation after semantic alignment. Target Purity measures resistance to speaker collapse, while Source Purity measures resistance to speaker fragmentation.}
  \label{tab:long-form-multispeaker}
  \footnotesize
  \textbf{(a) English to Chinese}
  \vspace{2pt}

  \begin{tabular}{lccc}
    \toprule
    System & Spk. SIM $\uparrow$ & Target Purity $\uparrow$ & Source Purity $\uparrow$ \\
    \midrule
    Qwen3.5-LiveTranslate & 0.47 & 0.39 & 1.00 \\
    Doubao AST 2.0 & 0.77 & 0.89 & 0.94 \\
    GPT Realtime Translate & 0.63 & 0.42 & 0.90 \\
    \textbf{X-Translator (ours)} & 0.77 & 0.85 & 0.93 \\
    \bottomrule
  \end{tabular}

  \vspace{6pt}
  \textbf{(b) Chinese to English}
  \vspace{2pt}

  \begin{tabular}{lccc}
    \toprule
    System & Spk. SIM $\uparrow$ & Target Purity $\uparrow$ & Source Purity $\uparrow$ \\
    \midrule
        Qwen3.5-LiveTranslate & 0.51 & 0.38 & 0.99 \\
    Doubao AST 2.0 & 0.78 & 0.91 & 0.90 \\
    GPT Realtime Translate & 0.64 & 0.51 & 0.74 \\
    \textbf{X-Translator (ours)} & 0.77 & 0.84 & 0.88 \\

    \bottomrule
  \end{tabular}
\end{table*}

X-Translator remains competitive when speakers alternate within a session (Table~\ref{tab:long-form-multispeaker}). For English-to-Chinese, it ties Doubao AST 2.0 for the highest segment-level Spk. SIM (0.77), while reaching 0.85 Target Purity and 0.93 Source Purity. For Chinese-to-English, its Spk. SIM is 0.77, only 0.01 below Doubao, with corresponding purities of 0.84 and 0.88. Doubao achieves the strongest overall purity scores (0.89/0.94 and 0.91/0.90), but X-Translator stays within 0.02--0.07 across the two directions.

The two purity measures expose failure modes that speaker similarity alone does not capture. Qwen3.5-LiveTranslate has near-perfect Source Purity (1.00 and 0.99) but low Target Purity (0.39 and 0.38), a pattern consistent with multiple source speakers collapsing into too few target-speaker clusters.  GPT Realtime Translate reduces this collapse relative to Qwen3.5-LiveTranslate, but shows greater fragmentation in Chinese-to-English, where Source Purity falls to 0.74. X-Translator's jointly high similarity and bidirectional purity indicate that its prompt routing usually preserves both speaker identity and the source-to-target speaker structure, although Doubao remains stronger on speaker-cluster consistency.

\subsection{Multilingual Speech Translation}

\begin{table}[t]
  \centering
  \caption{Multilingual speech translation results on FLEURS. Generated target speech is first transcribed with target-language ASR; COMET is computed from the resulting transcript for each language direction and then
  macro-averaged within each direction group. X denotes the other
  supported languages excluding Chinese and English.}
  \label{tab:multilingual}
  \footnotesize

  \begin{tabular}{lccc}
    \toprule
    Direction
    & \# Directions
    & COMET Avg. $\uparrow$
    & Std. $\downarrow$ \\
    \midrule
    Zh$\rightarrow$X & 19 & 0.68 & 0.06 \\
    X$\rightarrow$Zh & 16 & 0.69 & 0.04 \\
    En$\rightarrow$X & 19 & 0.69 & 0.06 \\
    X$\rightarrow$En & 16 & 0.70 & 0.09 \\
    \midrule
    \textbf{Overall} & 70 & 0.69 & 0.07 \\
    \bottomrule
  \end{tabular}
\end{table}

Table~\ref{tab:multilingual} summarizes 70 FLEURS translation directions supported by the evaluated ASR--MT--TTS backend intersection. The overall direction-macro-averaged COMET score is 0.69, and the four group averages occupy a narrow range from 0.68 to 0.70. Translation into English is highest on average (0.70), followed by translation into Chinese and from English (both 0.69), while translation from Chinese averages 0.68. This small range suggests that aggregate semantic quality is broadly similar across the four direction groups.

Variation among individual directions is nevertheless non-negligible. The standard deviation is largest for X$\rightarrow$En (0.09), despite that group's highest mean, and ranges from 0.04 to 0.06 for the other groups. The macro-average should therefore be read as evidence of broad backend coverage rather than uniformly strong performance for every language pair. These scores also measure the full cascade through target-speech ASR, so they combine errors from translation, synthesis, and transcription. Direction-level scores and coverage are retained in the released machine-readable evaluation artifacts; the target-language ASR setup and aggregation procedure are documented in the appendix.

\section{Limitations and Future Work}

X-Translator inherits the main limitations of cascaded S2ST systems. Errors from ASR can propagate to MT and TTS, and early segment commitment may remove context that later modules cannot recover. Because the components are not trained end to end, the system cannot jointly optimize translation quality, speaker similarity, prosody, and latency. The current design instead relies on explicit runtime policies for commitment and prompt routing, which makes the system controllable but also introduces policy-sensitive failure modes.

The speaker-aware mechanism is also limited by online speaker tracking and prompt quality. Overlapping speech, rapid turn-taking, noisy environments, and acoustically similar speakers can make speaker assignment ambiguous. Even when the correct speaker is selected, the prompt may be too short, too noisy, or stylistically mismatched with the current utterance. These limitations are especially important for long-form conversations, where early prompt errors can affect many later synthesized segments.

The comparison with proprietary systems is necessarily behavioral rather than mechanistic. Commercial APIs may use larger models, private data, internal caching, adaptive endpointing, or safety filters that are not visible to us. Therefore, our results should not be interpreted as a complete explanation of why a proprietary system succeeds or fails. Instead, the comparison positions X-Translator relative to systems that users can access, while X-Translator's internal logs provide the mechanistic analysis that black-box baselines cannot provide.

Future work will explore stronger commitment policies that combine lexical, acoustic, prosodic, and uncertainty cues. It will also study more robust speaker routing, better prompt quality control, and adaptive prompt selection between fixed and rolling prompts. Finally, we plan to expand evaluation to more language pairs, noisier real-world settings, longer conversations, and more diverse proprietary and open research baselines.

\section{Conclusion}

We presented X-Translator, a modular cascaded S2ST system for real-time, long-form, and multi-speaker speech translation. The system combines streaming ASR, MT, and prompt-conditioned TTS through a session-level controller that commits stable source segments and routes them to speaker-specific voice prompts. This design targets deployment constraints that are not fully captured by short-form sentence-level evaluation alone.

X-Translator establishes an open and controllable alternative to proprietary real-time translation systems. The current results illustrate the trade-offs of a low-cost cascaded architecture: X-Translator provides strong speaker preservation and low target-speech error rates in the evaluated short-form settings, but its latency and translation quality do not consistently surpass proprietary systems. By exposing segment commitment, prompt routing, and module-level timing, X-Translator provides a practical platform for studying how these system decisions shape deployment-oriented S2ST.
\bibliography{custom}

\appendix
\clearpage
\appendix
\onecolumn

\section{Long-Form Evaluation Protocol}

Table~\ref{tab:long-form-data-protocol} summarizes the single-speaker long-form protocol and the actual duration of the evaluated audio. We use ten recordings per dataset and feed each recording continuously to simulate streaming input. The samples are randomly selected from recordings with clear speech.

\begin{table}[H]
  \centering
  \caption{Single-speaker long-form evaluation protocol.}
  \label{tab:long-form-data-protocol}
  \footnotesize
  \begin{tabular}{lccccc}
    \toprule
    Dataset & Direction & \# recordings & Total (h:mm:ss) & Range (min:s) & Mean (min:s) \\
    \midrule
    TEDLIUM-3 & En$\rightarrow$Zh & 10 & 2:34:29 & 12:02--19:45 & 15:27 \\
    WenetSpeech & Zh$\rightarrow$En & 10 & 2:30:15 & 10:04--24:54 & 15:01 \\
    \bottomrule
  \end{tabular}
\end{table}

\section{Multi-Speaker Dataset Details}

Table~\ref{tab:multispeaker-data-details} documents the long-form multi-speaker evaluation sources and actual audio durations.

\begin{table}[H]
  \centering
  \caption{Multi-speaker evaluation data and durations.}
  \label{tab:multispeaker-data-details}
  \footnotesize
  \begin{tabular}{llcccp{0.28\textwidth}}
    \toprule
    Source & Direction & \# recordings & Total (h:mm:ss) & Range (min:s) & Usage notes \\
    \midrule
    VoxConverse & En$\rightarrow$Zh & 10 & 2:55:33 & 11:16--20:00 & Multi-speaker recordings evaluated under the dataset's research-use terms; audio is not redistributed \\
    Bilibili interviews & Zh$\rightarrow$En & 10 & 1:40:00 & 10:00--10:00 & Publicly accessible multi-speaker interviews are used only to produce research evaluation outputs; source audio is not redistributed \\
    \bottomrule
  \end{tabular}
\end{table}

Speaker prompts are transient excerpts derived from the evaluated source recording and are used only to condition synthesis within that evaluation session. We do not treat inferred speaker identities as labels and do not release source-derived voice prompts.

\section{Implementation and Runtime Configuration}

Table~\ref{tab:backend-runtime-config} reports the backend selection encoded by the released evaluation configuration. Unless explicitly noted, the same backend stack is used across the four evaluation settings.

\begin{table}[H]
  \centering
  \caption{Backend and runtime configuration.}
  \label{tab:backend-runtime-config}
  \footnotesize
  \begin{tabular}{lp{0.31\textwidth}p{0.22\textwidth}p{0.25\textwidth}}
    \toprule
    Module & Model/checkpoint & Serving mode & Key settings \\
    \midrule
    ASR & Qwen3-ASR & Local HTTP service & 16-kHz mono input; 1.6-s accumulated request chunk \\
    MT & LMT-60-8B & OpenAI-compatible local HTTP service & Maximum 256 generated tokens \\
    TTS & X-Voice (0.4B) & Local HTTP service & 48-kHz output; prompt-conditioned synthesis \\
    Speaker encoder & \texttt{iic/speech\_campplus\_sv\_zh-\allowbreak cn\_16k-common} & In-process ModelScope pipeline & 1.0-s embedding window with 0.5-s hop \\
    \bottomrule
  \end{tabular}
\end{table}

The released configuration connects the three generative modules through localhost endpoints and runs one multilingual language-pair worker by default. The implementation does not enable request batching or an application-level result cache. Hardware placement, numerical precision, and memory use are properties of the separately launched model servers rather than hard-coded X-Translator settings; we therefore do not infer them from the orchestration code.

\section{Multilingual Evaluation Details}

For every target language, generated speech is transcribed with target-language ASR before COMET is computed. The evaluation uses Whisper-medium with the target-language code supplied explicitly. Rather than repeating the same ASR configuration for every language, Table~\ref{tab:multilingual-asr-config} reports the common configuration once.

\begin{table}[H]
  \centering
  \caption{Target-language ASR configuration for multilingual evaluation.}
  \label{tab:multilingual-asr-config}
  \footnotesize
  \begin{tabular}{lp{0.28\textwidth}p{0.45\textwidth}}
    \toprule
    Targets & ASR model & Evaluation input and normalization \\
    \midrule
    All evaluated target languages & Whisper-medium & Generated target speech; the target-language code is set explicitly and the resulting transcript is passed to the shared OpenSTBench language-normalization policy \\
    \bottomrule
  \end{tabular}
\end{table}

Besides Chinese and English, the final input-language set contains 16 languages: Japanese, German, French, Portuguese, Italian, Spanish, Indonesian, Turkish, Vietnamese, Czech, Danish, Finnish, Malay, Norwegian Bokm\aa l, Polish, and Swedish. The final output-language set contains 19 languages besides Chinese and English: Japanese, German, Korean, Russian, French, Portuguese, Italian, Spanish, Indonesian, Thai, Turkish, Vietnamese, Czech, Danish, Finnish, Malay, Polish, Swedish, and Dutch. Consequently, Table~\ref{tab:multilingual} reports 19 directions from Chinese, 19 from English, 16 into Chinese, and 16 into English, for 70 evaluated cross-language directions in total. For each direction, samples are restricted to sentence IDs shared by its source and target FLEURS subsets; the source transcription and target-side transcription associated with the same sentence ID form the COMET source and reference, respectively.

We use \texttt{Unbabel/wmt22-comet-da}. COMET is computed from the ASR transcription of the generated target speech rather than from the MT module's direct text output. A sample is successful only when the system produces target audio and the target-ASR transcript is non-empty. We report COMET over successful samples and retain coverage separately; the evaluation artifacts also provide a conservative all-sample score in which failed samples receive zero. For a direction group $G$, the reported macro-average gives every direction equal weight:
\begin{equation}
  \operatorname{COMET}_{\mathrm{macro}}(G)
  = \frac{1}{|G|}\sum_{d\in G}\operatorname{COMET}(d).
\end{equation}
Same-language pairs are excluded. Direction-level scores and coverage are written by the released evaluator to machine-readable JSON and CSV files; we omit a long per-direction table from the paper appendix.

\section{Proprietary Baseline Configuration}

Table~\ref{tab:api-baseline-config} documents the proprietary systems as time-dependent behavioral baselines. Because service implementations can change, we identify the requested API model and the evaluation setting in which it was used rather than making claims about inaccessible internal checkpoints.

\begin{table}[H]
  \centering
  \caption{Configuration of proprietary speech translation baselines. Qwen3-LiveTranslate was used for the earlier short-form collection; Qwen3.5-LiveTranslate was used for the later long-form collection after the service was updated.}
  \label{tab:api-baseline-config}
  \footnotesize
  \begin{tabular}{lp{0.24\textwidth}p{0.18\textwidth}p{0.22\textwidth}p{0.20\textwidth}}
    \toprule
    System & Model/API identifier & Evaluation setting & Streaming configuration \\
    \midrule
    Qwen3-LiveTranslate & \texttt{qwen3-\allowbreak livetranslate-\allowbreak flash-\allowbreak realtime} & Short-form OpenSTBench & 16-kHz PCM over the DashScope realtime WebSocket API \\
    Qwen3.5-LiveTranslate & \texttt{qwen3.5-\allowbreak livetranslate-\allowbreak flash-\allowbreak realtime} & Long-form single- and multi-speaker & 16-kHz PCM over the DashScope realtime WebSocket API \\
    Doubao AST 2.0 & \texttt{doubao-ast-2.0} & Short- and long-form & 16-kHz PCM over the AST v2 WebSocket API in S2ST mode \\
    GPT Realtime Translate & \texttt{gpt-realtime-translate} & Short- and long-form & 24-kHz PCM over the Realtime Translations WebSocket API \\
    \bottomrule
  \end{tabular}
\end{table}

The short-form and long-form evaluations were conducted in separate collection periods. Qwen3-LiveTranslate was the evaluated service during the OpenSTBench collection; by the time the long-form experiments were run, the available service had been updated to Qwen3.5-LiveTranslate. This version difference reflects evaluation time rather than an experimental ablation, so comparisons across short- and long-form tables should not be interpreted as comparisons under an identical Qwen baseline.

Realtime clients use a 30-s connection timeout and a 180-s receive timeout by default, with up to three session-initialization attempts and a 2-s linearly increasing retry backoff. An output is successful when the service returns non-empty playable speech and a semantically complete translation under the OpenSTBench output checks. Failed calls are excluded from successful-only quality metrics and are reflected in the success rates reported below. For the short-form evaluation, each proprietary baseline is evaluated twice on the same sampled inputs, and Tables~\ref{tab:short-form-openstbench} and~\ref{tab:short-form-s2tt} report the arithmetic mean of the two run-level metric values.

\section{OpenSTBench API Success Rates}

Table~\ref{tab:openstbench-success-rate} reports the successful-call rates collected during the short-form OpenSTBench evaluation used in Table~\ref{tab:short-form-openstbench}. For the proprietary systems, an unsuccessful call includes a short input for which the system does not return a semantically complete translation. The lower success rates of Qwen3-LiveTranslate and Doubao AST 2.0 in some settings are mainly caused by this short-audio semantic-completeness filtering behavior rather than by a transport failure.

\begin{table}[H]
  \centering
  \caption{Successful-call rates on the short-form OpenSTBench evaluation by dataset and translation direction.}
  \label{tab:openstbench-success-rate}
  \footnotesize

  \begin{tabular}{lcccccc}
    \toprule
    System & mslt EN2ZH & mslt ZH2EN & speaker EN2ZH & speaker ZH2EN & ravdess & mcae\_spps \\
    \midrule
    Qwen3-LiveTranslate & 99.33\% & 97.00\% & 100.00\% & 100.00\% & 99.03\% & 99.90\% \\
    Doubao AST 2.0 & 93.50\% & 99.42\% & 100.00\% & 100.00\% & 100.00\% & 100.00\% \\
    GPT Realtime Translate & 94.25\% & 99.58\% & 100.00\% & 100.00\% & 100.00\% & 99.81\% \\
    X-Translator & 98.92\% & 100.00\% & 100.00\% & 100.00\% & 99.97\% & 100.00\% \\
    \bottomrule
  \end{tabular}
  
\end{table}

\end{document}